\documentclass{amsart}

\numberwithin{equation}{section}
\usepackage{graphicx}
\usepackage{hyperref}
\begin{document}

\title[]{``Violating'' Clauser-Horne inequalities within classical mechanics}
\author{S. Antoci \and L. Mihich}
\address{ Dipartimento di Fisica ``A. Volta'' and I. N. F. M.,
Via Bassi 6, Pavia, Italy}
\email{Antoci@fisav.unipv.it}

\begin{abstract}
Some authors have raised the question whether the probabilities
stemming from a quantum mechanical computation are entitled to
enter the Bell and the Clauser-Horne inequalities. They have
remarked that if the quantum probabilities are given the status of
conditional ones and the statistics for the various settings of
the detectors in a given experiment is properly kept into account,
the inequalities happen to be no longer violated. In the present
paper a classical simile modeled after the quantum mechanical
instances is closely scrutinised. It is shown that the neglect of
the conditional character of the probabilities in the classical
model leads not only to ``violate'' the Clauser-Horne
inequalities, but also to contradict the very axioms of classical
probability theory.
\end{abstract}
\maketitle

\section{Introduction}
Many years have elapsed since Bell \cite{ref:Bell65},
\cite{ref:Bell71} nailed down his prolegomena to any future
hidden-variables theory, and Horne and Clauser worked out their
inequalities \cite{ref:CH}, that shifted the whole issue from the
realm of the {\it Gedankenexperimente} to the optical
laboratories. And from these workshops of empirical evidence the
response came, more and more convincing with the lapse of the
years and the consequent refinement of the techniques for
experimentation, aimed at closing all the possible loopholes
\cite{ref:Clauser76}, \cite{ref:Aspect81}, \cite{ref:Zeilinger98}.
Needless to say, the experiments without exceptions have
confirmed, although not to everybody's satisfaction, the
predictions stemming from quantum mechanics. Had not this been the
case, theoretical physicists would have found something very
interesting to do during the last years. But alas, quantum
mechanics works finely even with these pairs of photons, whatever
a photon may turn out to be \footnote{ In a letter of December 12,
1951 Albert Einstein wrote to his long-time friend Michele Besso:
``The whole fifty years of conscious brooding have not brought me
nearer to the answer to the question `What are light quanta?'.
Nowadays every scalawag believes that he knows what they are, but
he deceives himself.''\cite{ref:Speziali72} This English
translation of the excerpt can be found in an article by J.
Stachel \cite{ref:Stachel86}.}. Under this respect, the above
mentioned experiments did not have much to add to what was already
common wisdom for the overwhelming majority of physicists since
many decades: quantum mechanics is the theory of choice for
describing the microscopic reality in its interaction with
macroscopic detectors.\par But of course this is only one aspect
of the question. The reason why the outcome of such delicate
experiences is scrutinised by the theoreticians with so much
attention depends also on their epistemological charme: not only
is quantum mechanics once more confirmed by the experiments, but
also (apart from residual loopholes) the so called local
hidden-variables theories appear deprived of any residual hope for
challenging quantum mechanics as the sole ruler of the microscopic
realm \cite{ref:CH}. At first sight, one does not grasp wherefrom
the pressure for settling so abstract a question could come, for
no credible pretenders to the role presently kept by quantum
theory have emerged up to now. For this reason the attempt at
outlining {\it a priori} the features of such would be pretenders
is fraught with the unavoidable dangers of vagueness. Despite
these risks, much effort has been spent for endowing the phantom
pretenders with formal attributes that could allow for more and
more general arguments against them. Much less work has been done
(with some notable exceptions, like the ones accounted for in
\cite{ref:Marshall97}, \cite{ref:Galgani99}) for understanding
what the good old classical physics \footnote{``{\it Die Physik
der Modelle}'', as Schr\"odinger nostalgically dubbed it
\cite{ref:Schr35}.} might have to suggest, through the analysis of
particular instances, as hypothetical patterns of behaviour.

\section{The Clauser-Horne inequalities}
With the optical experiment in mind, Clauser and Horne have
envisaged the behaviour that a wide class of stochastic local
theories, designated by them as {\it objective local theories},
would display in the considered experimental circumstances. For
completeness we summarise here the features of the model situation
contemplated in \cite{ref:CH}. A common source emits pairs of
correlated entities that fly or propagate in opposite directions.
When one of them reaches a detector, a characteristic feature of
the entity, say the way it lies in a plane normal to the direction
of motion, can be measured. Since the putative theory has to be an
objective one, the entity is supposed to possess this feature in a
way independent of any act of measurement. The overall behaviour
of the single pair is characterized by a ``hidden'' parameter
$\lambda$ in the following sense: $\lambda$ would specify the
initial state of the pair and its subsequent development when the
mutual interaction of the components has stopped, were it not for
a residual element of uncertainty. The authors of \cite{ref:CH} do
not specify the nature of this uncertainty \footnote{The relation
of equivalence between deterministic and stochastic
hidden-variables models was first elucidated \cite{ref:Fine82} by
A. Fine.}. Anyway, a residual randomness \cite{ref:Synge83} is
left, maybe intrinsic, maybe just due to our laziness in
investigating the system. Therefore it makes sense to speak of the
conditional probability $p(A|\lambda)$ that the entity that
propagates, say, to the left, when reaching the detector set on
the left will display there a characteristic feature, for instance
an angle $A$, when the hidden parameter has the value $\lambda$.
Let $p(B|\lambda)$ be the conditional probability that the entity
going to the right be detected to display as characteristic
feature an angle $B$ when the hidden parameter has the value
$\lambda$. Clauser and Horne make an assumption, which they claim
to be ``a natural expression of a field-theoretical point of view,
which in turn is an extrapolation from the common-sense view that
there is no action at a distance''. They stipulate that the local
objective theories should obey the following factorisability
condition:
\begin{equation}\label{2.1}
p(A,B|\lambda)=p(A|\lambda)p(B|\lambda)
\end{equation}
for the joint probability $p(A,B|\lambda)$ of detecting the left
entity as displaying the angle $A$ and the right entity as
displaying the angle $B$ when the hidden parameter has the value
$\lambda$. Let $A$, $A'$, and $B$, $B'$ be the four angles that
happen to be displayed by the propagating entities at their
respective acts of detection when the hidden parameter has the
value $\lambda$. It is an easy matter \cite{ref:CH} to show that
the inequalities
\begin{eqnarray}\label{2.2}
-1\leq p(A|\lambda)p(B|\lambda)-p(A|\lambda)p(B'|\lambda)
+p(A'|\lambda)p(B|\lambda)+p(A'|\lambda)p(B'|\lambda)
\nonumber\\-p(A'|\lambda)-p(B|\lambda)\leq 0
\end{eqnarray}
must hold. By integrating $p(A|\lambda)$ and $p(A,B|\lambda)$ over
$\lambda$ with a suitably normalised weight function
$\rho(\lambda)$ one defines the probability
\begin{equation}\label{2.3}
p(A)\equiv\int\rho(\lambda)p(A|\lambda)d\lambda,
\end{equation}
the joint probability
\begin{equation}\label{2.4}
p(A,B)\equiv\int\rho(\lambda)p(A|\lambda)p(B|\lambda)d\lambda,
\end{equation}
and eventually finds the now famous Clauser-Horne inequalities
\begin{eqnarray}\label{2.5}
-1\leq p(A,B)-p(A,B')+p(A',B)+p(A',B')-p(A')-p(B)\leq 0
\end{eqnarray}
as a necessary condition that all the objective local theories
must fulfil. One cannot help agreeing with Clauser and Horne that
the factorisability condition (\ref{2.1}) is a necessary one in
the objective local theories; one can add that, apart from the
above mentioned exceptions \cite{ref:Marshall97},
\cite{ref:Galgani99}, {\it die Physik der Modelle} generally
requires the satisfaction of such a condition, hence the general
validity of the Clauser-Horne inequalities. When the quantum
mechanical probabilities, let us say $q(A)$, $q(A,B)$, etc.,
calculated for particular instances dealing either with correlated
spins or with correlated light quanta are substituted for the
corresponding probabilities in (\ref{2.5}), it is found that, if
the angles are appropriately chosen, the resulting expression does
not fulfil the inequalities. According to the established wisdom,
this is the hallmark of quantum mechanics, well confirmed by the
experimental facts (apart from residual loopholes), a unique
feature that cannot be mimicked by classical physics, unless one
contemplates either subtle enhancement processes connected {\it
e.g.} with the existence of a zero-point electromagnetic field
\cite{ref:Marshall97} or a nonlocal behaviour, like the one
stemming from the Lorentz-Abraham-Dirac equation of motion for the
classical electron \cite{ref:Galgani99}.\par In recent years,
however, rather lonely but persistent voices have been heard
\cite{ref:Brans88}, \cite{ref:Szabo95a}, \cite{ref:Szabo95b},
\cite{ref:Durt96}, \cite{ref:Bana97}, claiming {\it inter alia}
that some not so venial sin is committed when substituting the
quantum mechanical probabilities for the probabilities $p(A)$,
$p(A,B)$ in (\ref{2.5}), and that when the wrongdoing is amended,
the Clauser-Horne inequalities happen to be no longer violated.
More than seven decades have elapsed since the very notion of
quantum probabilities has been hesitantly extracted
\cite{ref:Born26a}, \cite{ref:Born26b} by Born from Einstein's
idea of the {\it Gespensterfeld}, and yet one feels some
uneasiness when forced to confront a seemingly simple question
like this. One would venture in this sort of discussion with a
lighter heart, had the number of interpretations attached since
then to quantum mechanics shrinked instead of increasing, as it
has been the case, and provided that the quantum measurement
problem had already been solved for good, {\it i.e. more
relativistico}. While waiting for such occurrences, one still
feels to walk on firmer ground if once more this admittedly old
fashioned and just preliminary approach is attempted: seeking
whether a more familiar simile, entirely grounded on classical
physics and on classical probability theory, may provide some
enlightenment.

\section{A classical model that ``violates'' the inequalities}
Let us consider an experimental device like the one drawn in Fig.
1, whose working is entirely accounted for by classical mechanics:
two equal cylindrical bodies, ``1'' and ``2'', are thrown from the
spatial origin of an inertial reference frame with opposite
velocities $v_1=-v_2$ and with opposite angular velocities
$\omega_1=-\omega_2$, all directed along the $y$ axis, by the
action of, say, a pressed and twisted spring interposed between
them and suddenly released. The action of the spring lasts for a
very short time, after which the two material bodies are free to
run and to turn around their axes in the interior of a hollow
cylinder of length $l$, kept at rest in the considered inertial
reference system. The hollow cylinder is centered at the origin of
the spatial coordinates and its axis lies in the $y$ direction;
its scope is to act as measuring device. To this end on its left
half, along the whole span $-l/2<y<0$, two straight marks are
engraved, whose projections on the $x,z$-plane happen to lay at
the angles $A$ and $A'$ with respect to the $x$ axis. Two straight
marks are engraved also on the right half of the hollow cylinder,
along its whole span $0<y<l/2$, at the angles $B$ and $B'$
respectively. On the inner rims of the cylindrical bodies ``1''
and ``2'' two tiny marks $m_1$ and $m_2$ are impressed, and we
shall abide to the rule that, after the pressing and twisting of
the spring is accomplished, the two marks shall happen to coincide
at some angle $\varphi$ measured as previously in the $x,z$-plane.
We shall also take care to load the spring always in the same way;
the rotation angles of the two bodies, when they run from the
center to the ends of the measuring device, shall be invariably
$\gamma_1=const.>0$ for the body running to the left, and
$\gamma_2=-\gamma_1$ for the other one.\par
\begin{figure}[h]
\includegraphics{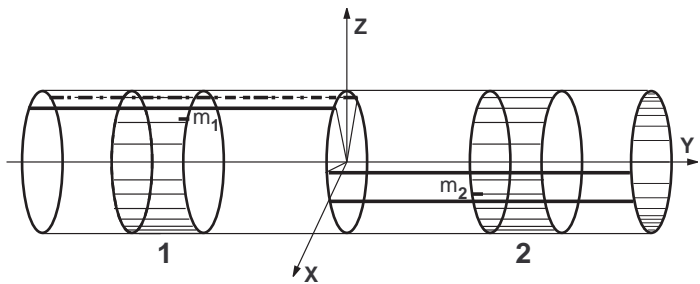}
\caption{Schematic rendering of the experimental device
aimed at ``violating'' the Clauser-Horne inequalities within
classical mechanics. The heavy lines correspond to the marks
engraved on the outer cylinder, respectively at the angles $A$, $A'$ of
its left half and at the angles $B$, $B'$ of its right half. The
spring, the mechanical constraint linking the bodies ``1'' and
``2'' and the pair of mechanical stops set at the angles, say,
$A$ and $B$ are left to the imagination of the reader.}
\end{figure}
\noindent Let us indulge, with the given apparatus, in the following
exercise of ``experimental probability''. We pick at random
some angle $\varphi$ from a uniform distribution that
extends between $0$ and $2\pi$, we load
and twist the spring interposed between the bodies as prescribed
above, and we eventually set the coincident marks $m_1$ and $m_2$
at the angle $\varphi$ before releasing the spring. We then check
whether the mark on body ``1'', that runs to the left, crosses or
not the straight lines drawn at the angles $A$ and $A'$, and
whether the mark on body ``2'' reaches or not the angles $B$ and
$B'$. After repeating {\it ab ovo} the whole procedure $N$ times,
the statistics of the experiment can be compiled, and the
``experimental probability'' can be eventually inferred. Of course
nobody will really indulge in so trivial an experiment, since the
physical situation is quite clear, and the classical probabilities
$p(A)$, $p(A,B)$, etc. can be directly evaluated through a very
simple geometric argument. Needless to say, when these
probabilities are inserted in the Clauser-Horne inequalities
(\ref{2.5}), the latter happen to be fulfilled for all the
possible choices of the angles $A$, $A'$, and $B$, $B'$. Suppose
however that we are caught by a virulent form of wishful thinking:
we would like to see these inequalities violated, despite the fact
that the instruments at our disposal are constituted by purely
classical, objective and local entities. Therefore, in order to
mimic the quantum mechanical behaviour, we modify the experimental
device described above, and we conjure up an illusive
``Verschr\"ankung'' by:
\begin{itemize}
\item connecting the bodies ``1'' and ``2'' through a mechanical
constraint, that hinders their rotations when the mutual rotation
angle reaches the value $\gamma$, in order to provide a not quite
mysterious surrogate to the ``spooky action at a distance'',
\item allowing for an active role of the measuring device, through
mechanical stops placed in the interior of the hollow cylinder,
respectively at an angle, say $A$, on the left side, acting in the
span $-l/2<y<0$, and at an angle $B$ on the right side, acting in
the span $0<y<l/2$. These stops are so contrived as to block the
motion of the bodies ``1'' and ``2'' whenever the tiny marks $m_1$
and $m_2$ impressed on their inner rims respectively reach the
above mentioned angles. It is intended that we can change the
position of the mechanical stops from the pair of angles $A,B$ to
the pairs $A,B'$, $A',B$ and $A',B'$, or even suppress one of the
stops, while leaving the other one active and set just at one of
the four positions mentioned above.
\end{itemize}

\noindent We test the modified device by choosing the angles as
shown in Figure 2, {\it i.e.} we set $B'=0$, $B=\vartheta$,
$A'=\gamma$, $A=\gamma+\vartheta$, with $0<\vartheta<\gamma$; the
spring is loaded just in the way kept earlier, that entails a
clockwise rotation of the body ``2". It is our intention to perform
again a sequel of trials with the start angle $\varphi$
picked at random in a uniform distribution between $0$ and $2\pi$,
as it was appropriate with the earlier version of the apparatus,
but our plan meets with a certain difficulty: while previously
just one series of trials, corresponding to just one experimental
setup, was sufficient for inferring all the ``experimental
probabilities'', the situation now is completely different. In
order to gather the statistics required for inferring the joint
probabilities four distinct sequences of trials are needed, one
for each of the physically different setups that can occur when
both mechanical stops are active. Furthermore, in order to infer
the probabilities of the single occurrences, four additional
sequences of trials seem needed, one for each of the physically
different situations that can occur when one of the mechanical
stops is removed. What we infer from the statistics of these eight
independent sequences of trials, performed each one on a different
physical system, are of course {\it conditional} probabilities.
They are the probabilities for the bodies to reach certain angles
when the stops are placed in a certain way. Also in this case there
is no need to perform really the sequences of trials.
\begin{figure}[t]
\includegraphics{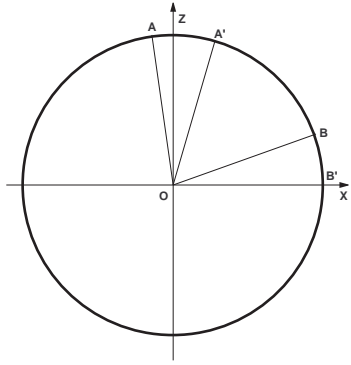}
\caption{Representation on the trigonometric circle of the
$x,z$-plane of the angles $B'=0$, $B=\vartheta$, $A'=\gamma$,
$A=\gamma+\vartheta$. They give the possible positions of the mechanical
stops that allow for the ``violation'' of the Clauser-Horne inequalities
described in the text.}
\end{figure}
Let $p(A,B|a,b)$ mean the conditional probability that the
marks on the two bodies reach the angles $A$ and $B$ when the mechanical
stops are both active and set just at the angles $a=A$, $b=B$,
while $p(A|a)$ means the conditional probability that the mark on
body ``1'' reach the angle $A$ when only the mechanical stop on
the left is active, and set at the angle $a=A$. Simple geometric
arguments give
\begin{equation}\label{3.1}
p(A,B|a,b)=p(A',B'|a',b')={\gamma\over{2\pi}},~~~
p(A,B'|a,b')=0,~~~p(A',B|a',b)={{\gamma-\vartheta}\over{2\pi}},
\end{equation}
while
\begin{equation}\label{3.2}
p(A|a)=p(A'|a')=p(B|b)=p(B'|b')={\gamma\over{4\pi}}.
\end{equation}
It is quite clear that we should abstain from the inconsiderate
act of plugging these quantities in the Clauser-Horne inequalities
(\ref{2.5}), since each one of them is a conditional probability
referring to a different physical system, while the inequalities
deal with the probabilites for just one physical system. If we
insist in doing so, a nonsensical outcome must be expected, and is
in fact obtained. The inequality (\ref{2.5}) is ``violated'' on
the right, despite the fact that classical mechanics fully
accounts in an objective and local way for all the physical
happenigs considered here. Moreover, through the same hocus-pocus
also the inequality
\begin{eqnarray}\label{3.3}
-1\leq p(A',B')-p(A',B)+p(A,B')+p(A,B)-p(A)-p(B')\leq 0,
\end{eqnarray}
that can be obtained from (\ref{2.5}) by exchanging the angles
with and without a prime, is ``violated'' on the right. By summing
the inequalities stemming from the two ``violations'' one would
get
\begin{equation}\label{3.4}
2p(A,B)+2p(A',B')-p(A)-p(A')-p(B)-p(B')>0.
\end{equation}
Of course one expects that in an entirely classical situation at
least Bayes' axiom
\begin{equation}\label{3.5}
p(A,B)=p(A)p(B|A)=p(B)p(A|B)
\end{equation}
should be respected. But if we insist on this requirement and
insert (\ref{3.5}) in (\ref{3.4}), we eventually reach an
unquestionably absurd result: at least one of the four conditional
probabilities $p(A|B)$, $p(B|A)$, $p(A'|B')$, $p(B'|A')$ should be
$>1$! By inserting the conditional probabilities (\ref{3.1}) and
(\ref{3.2}) in the Clauser-Horne inequalities we have just
committed a blunder. But the way out is not so arduous, and we
could have spared the useless trials with one stop removed. The
four trials with the two stops active are in fact sufficient for
mastering the problem, if we confront it in the correct way. The
measuring device, after the addition of the mechanical stops, is
no longer a passive entity. Due to their presence the physical
system under investigation is not constituted only by the bodies
``1'' and ``2'' and by the mechanical constraint that hinders
mutual rotations of the latter larger than the angle $\gamma$. The
physical system now includes the measuring device with its
mechanical stops, and the probabilities that enter the
Clauser-Horne inequalities shall deal with this larger system. Let
us discard the trials that were performed with only one stop
active, and consider the ratios between the number of trials
performed with a certain setting of the two stops and the overall
number of trials performed with the two stops both active. From
these ratios we infer, in retrospect, the probabilities for the
various settings of the stops. We call them $p(a,b)$, $p(a,b')$,
$p(a',b)$, $p(a',b')$. Then the probabilities concerning the whole
experiment, that we are entitled to insert in the Clauser-Horne
inequalities, shall read:
\begin{eqnarray}\label{3.6}
p(A,B)=p(A,B|a,b)p(a,b),~~~p(A,B')=p(A,B'|a,b')p(a,b'),\nonumber\\
p(A',B)=p(A',B|a',b)p(a',b),~~~p(A',B')=p(A',B'|a',b')p(a',b'),
\nonumber\\
p(A)=p(A,B)+p(A,B'),~~~p(A')=p(A',B)+p(A',B'),\nonumber\\
p(B)=p(A,B)+p(A',B),~~~p(B')=p(A,B')+p(A',B').
\end{eqnarray}
By inserting (\ref{3.6}) in (\ref{2.5}) the latter reduces to
\begin{equation}\label{3.7}
-1\leq -p(A,B'|a,b')p(a,b')-p(A',B|a',b)p(a',b)\leq 0
\end{equation}
which is of course not violated {\it in general} both on its right
and on its left side.

\section{Conclusion}

We ask for the reader's forgiveness, because the scrutiny of the
detailed working of our artful ``Verschr\"ankung'' may well have
looked a pedantic recitation of things so well known that their
further recollection is just a waste of time. But what is given
for granted in classical physics does not seem to be so well
settled in quantum physics if, after so many years since the
appearance \cite{ref:CH} of the Clauser-Horne inequalities and
after so many papers written on the subject, some authors
\cite{ref:Szabo95a}, \cite{ref:Szabo95b}, \cite{ref:Durt96} still
felt urged to stand up and to remark that a major conceptual
error, akin to the one expounded in the previous Section, has been
committed in quantum physics. They have shown also that if the
quantum probabilities are dealt with as {\it conditional}
probabilities and the statistics of the use of the detectors is
taken into account, quantum mechanics no longer happens to violate
the inequalities in the situations that have been so carefully
scrutinised, both by theoreticians and by experimentalists. The
readings of the above mentioned papers and of the seminal work
\cite{ref:Brans88} by C. Brans have convinced us that even a
recollection of the obvious, as it has been done with the
classical simile of the present paper, could be helpful in
intimating what might be the proper use of the Clauser-Horne
inequalities in quantum mechanics.

\bibliographystyle{amsplain}

\end{document}